\documentstyle[aps,pra,twocolumn,psfig]{revtex}
\def\be{\begin{equation}}
\def\ee{\end{equation}}
\def\bea{\begin{eqnarray}}
\def\eea{\end{eqnarray}}
\author{Dominique Delande$^{1}$ and Jakub Zakrzewski$^{2}$}
\address{
$^1$ Laboratoire Kastler-Brossel, \\
Tour 12, Etage 1, 4 Place Jussieu, F-75005 Paris,\\
$^2$ Instytut Fizyki imienia Mariana Smoluchowskiego, Uniwersytet
 Jagiello\'nski,\\ ulica Reymonta 4, PL-30-059 Krak\'ow.
}
\title{Spontaneous emission of non-dispersive Rydberg wave packets}
\date{\today}

\begin{document}

\maketitle

\begin{abstract}
Non dispersive electronic Rydberg wave packets may be created in atoms
illuminated by a microwave field of circular polarization.
 We discuss the spontaneous emission
  from such states and 
show that the elastic incoherent component
(occuring at the frequency of the driving field) dominates
the spectrum in the semiclassical limit, contrary to earlier predictions.
We calculate
the frequencies of single photon emissions and the associated rates
in the ``harmonic approximation",
i.e. when the wave packet has approximately a Gaussian shape.
The results 
agree well with exact quantum mechanical calculations, which validates
the analytical approach.
\end{abstract}

\draft
\narrowtext

\section{Introduction}

Recent theoretical studies predicted the existence
of non-dispersive wave packets in atoms driven by a microwave radiation
of either circular \cite{bke} or linear polarization \cite{ab2}.
Contrary to standard electronic wave packets which disperse in time
when the binding force is nonlinear, these novel objects  do not
spread due to a nonlinear
coupling between the atom and the driving field
which locks the electronic motion onto the driving frequency.
The phenomenon responsible for this locking is a 1:1 resonance
between the internal (Kepler) and external (microwave) frequencies,
creating a resonance island at the center of which lies a stable
periodic orbit. 
In the dressed atom picture \cite{cct},
a single eigenstate
of the full ``dressed" hamiltonian represents
the wave packet \cite{ab2,dzb95}, explaining immediately 
the absence of spreading. 

The very same microwave field is also responsible for the decay of
the wave packet via ionization. As shown elsewhere \cite{zdb95},
the typical ionization time of wave packets built from circular
Rydberg states around principal quantum number $n_0=60$ exceeds $10^5$
Kepler periods being often longer than $10^6$ periods. The ionization
occurs via a ``chaos assisted tunnelling'' mechanism \cite{zdb98} (the
tunneling occurring from the stable island into the chaotic sea surrounding
it). In effect the lifetime against ionization is very sensitive
to small changes of the parameters (microwave frequency, $\Omega$ or its
amplitude $F$) exhibiting large scale fluctuations \cite{zdb98}.

Another possible decay mechanism of the wave packet is of course the
spontaneous emission of photons. 
For circularly polarized microwave, the situation considered in this paper,
it has been discussed in \cite{bb97}.
Here we readdress this issue 
since, in our opinion, the results reported in \cite{bb97} are
incomplete. To this end, we use a  different approach, borrowing
heavily from the well known treatment of resonance fluorescence of
two-level atoms driven by a laser field \cite{cct} using the dressed atom
approach. 
The main advantage of the present approach is its generality: it can
be used for any polarization of the microwave field, for any
superposition of driving fields with various frequencies and for
any structure of electromagnetic modes surrounding the atom 
(as shown e.g. in \cite{vacuum}, the presence of a
cavity can strongly affect the spontaneous emission of an atom).
Also, it uses the common knowledge widely developed in the 
quantum optics community and is thus accessible to a large audience.
Let us mention also that spontaneous emission of 
non-dispersive wave packets in
linearly polarized microwave is discussed in \cite{ab97}.

The paper is organized as follows. In section II, we discuss
the basic properties of the non-dispersive wave packets 
(not considering spontaneous emission) from different
points of view: laboratory frame or rotating frame, time-dependent
or Floquet pictures, dressed atom picture. The discussion is
necessary for a proper understanding of spontaneous emission,
which is done in section III. Our results are compared to the
earlier published results in section IV. Section V is devoted to
the validity of the harmonic approximation, by comparing
our analytical results to ``exact'' numerical calculations.
We conclude in section VI.

\section{The non-dispersive wave packets from different points of view}

We consider a hydrogen atom in the presence of a 
monochromatic driving microwave field
of circular polarization, that is an electromagnetic field
which can be described by a single mode. In this section, we will {\em not}
consider any coupling to other (initially empty) modes of the
electromagnetic field. This will be discussed in section III.
We first treat the microwave field classically.
A full quantum treatment of the microwave field is possible
(see section \ref{dressed}), but not really needed as, for usual microwave
fields, the number of photons in the mode is enormous and a classical
treatment of the microwave field is possible.

\subsection{Hamiltonian in the laboratory frame}

With the previous assumption, the quantum Hamiltonian of the
system is
(in atomic units, neglecting relativistic and QED effects, assuming
an infinite mass of the nucleus and in the dipole approximation):
\begin{equation}
H_{\rm cl}= \frac{{\bf p}^2}{2} -\frac{1}{r} +F(x
\cos\Omega t +y \sin\Omega t).
\label{hcl}
\end{equation}
where $\Omega$ is the frequency of the microwave field circularly
polarized in the $x-y$ plane and $F$ its amplitude. The $x$-axis
is chosen to coincide with the microwave electric field at $t=0.$

\subsection{Floquet approach}

\label{floquet}

The Hamiltonian (\ref{hcl}) is periodic in time with period $2\pi/\Omega.$
Hence, a 
Floquet approach \cite{gf} may be
used: any solution of the time-dependent Schr\"{o}dinger
equation can be expanded as a linear combination of {\em time-periodic}
states - called Floquet states of the system - with coefficients 
oscillating in time as $\exp (-iE_it)$ where $E_i$ is called the
quasi-energy of the Floquet state (there is no explicit
$\hbar$ in the equations as we are using atomic units where it is one). 
Obviously, the quasi-energy is defined
only modulo $\Omega$ making the Floquet spectrum periodic
in energy with period precisely $\Omega.$
The Floquet spectrum as well as the Floquet states can be obtained
from the diagonalization of a suitable operator, namely the 
Floquet Hamiltonian:
\begin{equation}
{\cal H}= H_{\rm cl} - i \frac{\partial}{\partial t}
\label{hamfloq}
\end{equation}
with time-periodic boundary conditions.
A suitable basis for the Floquet Hamiltonian is the product of
any basis in usual space - for example, the usual hydrogenic 
eigenbasis with bound states $|nlm\rangle$ - by the basis of
time-periodic oscillating exponentials
$\exp (iK\Omega t)$ with $K$ an integer. We will denote
$|nlm;K\rangle$ such a basis.
The next step is to expand the time-periodic Hamiltonian in a Fourier
series and calculate the non-zero matrix elements of the Floquet Hamiltonian.
Because no 
higher harmonics of the microwave frequency exist in the Hamiltonian,
only matrix elements changing $K$ by zero or one unit do not vanish.
Moreover, in the specific case of circular polarization, the part in
the Hamiltonian which increases (resp. decreases) $K$ by one is 
proportional to $x-iy$ (resp. $x+iy$) hence decreases (resp. increases)
the angular momentum of the electron along the $z$ axis by one unit.
In other words, the Floquet Hamiltonian obeys the 
selection rule $\Delta (K+m) = 0.$
This signifies
merely that absorption of a $\sigma^+$ polarized microwave photon
must increase the atomic angular momentum $L_z=xp_y-yp_x$ by one unit.
The Floquet Hamiltonian thus has a block diagonal structure and
the Hilbert space can be split into manifolds characterized by
constant value of ${\cal K}=m+K.$ In each block, the Floquet Hamiltonian
is no longer periodic in energy. On the other hand, the Floquet
Hamiltonian in blocks ${\cal K}$ and ${\cal K}'$ only differ
by a constant energy term $({\cal K}'-{\cal K}) \Omega$
which restores the total periodicity of the Floquet spectrum.
However, diagonalization in a given block is sufficient to obtain
all the information on the system.

\subsection{Hamiltonian in the rotating frame}

For a circularly polarized radiation, another possibility exists which
has been frequently used in the past 
\cite{bke,dzb95,zdb95,zdb98,bb97,keb,ke,bb97r,bb96,flul,lbfl,flu,buf,lbf,clfu,dzb97}.
By passing to the non inertial frame rotating with frequency
$\Omega$ using the unitary transformation $U=\exp(i\Omega L_z t),$
one may remove the oscillatory time dependence from Eq.~(\ref{hcl})
obtaining in the rotating frame the Hamiltonian
\begin{equation}
H_{\rm rot}= U \ H_{\rm cl}\ U^{\dagger} + 
i U \frac{\partial U^{\dagger}}{\partial t}
=
\frac{{\bf p}^2}{2} -\frac{1}{r} +Fx -\Omega L_z. 
\label{hrot}
\end{equation}

This time-independent Hamiltonian has some energy levels and corresponding
eigenstates. Obviously, its spectrum is not $\Omega$-periodic.
However, the link with the Floquet approach in the laboratory
frame is easily established: there is a one-to-one correspondence
between the eigenstates in the rotating frame and the Floquet states in the
laboratory frame from the block ${\cal K}=0.$ Indeed, consider an eigenstate
(i.e. time-independent  state)
$|\phi_i\rangle$ of $H_{\rm rot}$ with energy $E_i.$ From Eq.~(\ref{hrot}),
we deduce straightforwardly: 
\begin{equation}
\left(H_{\rm cl} -i \frac{\partial}{\partial t} \right) 
U^{\dagger}|\phi_i \rangle = E_i U^{\dagger}|\phi_i \rangle
\end{equation}
which implies (see Eq.~(\ref{hamfloq})) that $U^{\dagger}|\phi_i \rangle$ is
a Floquet eigenstate with quasi-energy $E_i.$ From the definition
$U^{\dagger} = \exp(-i\Omega L_z t),$ it is clear that the $K$-th Fourier
component of $U^{\dagger}|\phi_i \rangle$ is nothing but the component of
$|\phi_i \rangle$ on the subspace with $L_z=m=-K.$ In other words, 
$U^{\dagger}|\phi_i \rangle$ is a Floquet eigenstate in the laboratory
frame lying in the block ${\cal K}=m+K=0.$

Of course, Floquet states belonging to the other blocks ${\cal K}\ne 0$
can also be found from the time-independent Hamiltonian in rotating frame.
A first method is to use, instead of the operator $U$, a slightly different
unitary equivalent operator $\exp (-ik\Omega t)U$ (with $k$ an integer, it has
the same time-periodicity 
 as $U$ itself)
which leads to the following
Hamiltonian in the rotating frame:
\begin{equation}
H_{\rm rot}=
 \frac{{\bf p}^2}{2} -\frac{1}{r} +Fx -\Omega (L_z-k). 
\label{hrot1}
\end{equation}
Taking all possible positive and negative integer $k$ values
one recovers the full Floquet spectrum.

A second equivalent method is to keep the standard Hamiltonian in rotating frame,
Eq.~(\ref{hrot}), and to consider states of the form 
$\exp (ik\Omega t)U^{\dagger}|\phi_i \rangle$. These are time-periodic Floquet
(if $k$ is an integer), eigenstates of the Floquet Hamiltonian ${\cal H}$,
Eq.~(\ref{hamfloq}) with energy $E_i+k\Omega$, belonging to the block
${\cal K}=m+K=k.$

At a deeper level, this can be seen as a consequence of the unitary equivalence
(which implies the same spectrum) for the Floquet Hamiltonians in the laboratory frame
and in the rotating frame. The spectrum of the latter is here very simple: 
because $H_{\rm rot}$ is time-independent, it is simply the energy spectrum
of $H_{\rm rot}$ shifted by an arbitrary integer multiple of $\Omega.$
For a more complicated case like -- for example -- a microwave with polarization
close to circular, the Hamiltonian in the rotating frame is weakly
time-dependent and its corresponding Floquet Hamiltonian (with the time dependent
term treated perturbatively) may be the simplest approach.

\subsection{Dressed atom approach}

\label{dressed}

An alternative view of this problem is to treat the microwave field
quantum-mechanically as being
in either a coherent state or a pure Fock state with a large
number of photons in the circularly polarized mode \cite{foot1}.
This approach is quite common in quantum optics and is known as
the dressed atom picture. 
Then the atom is not considered as an isolated
quantum system, but it is rather the strongly coupled atom+field
system which is treated using quantum mechanics.
The Hamiltonian of the atom+field system is:
\begin{equation}
H_{\rm q}= \frac{{\bf p}^2}{2} -\frac{1}{r} + V (d_-a^\dagger + d_+a)
+\Omega a^\dagger a,
\label{hq}
\end{equation}
where $a^\dagger/a$ are the creation/annihilation operators in the 
occupied, $\sigma^+$ circularly
polarized mode, $d_\pm=(x\pm iy)/\sqrt{2}$ 
the corresponding components of the dipole
moment operator and $V$ is the coupling strength, whose exact expression
is given in section \ref{spont} but not important at the present level.

The Hamiltonian $H_{\rm q}$ is time independent and may
be expanded in the basis: $\{ |nlm\rangle \otimes |N\rangle =
|nlm;N\rangle \}$ where $|nlm\rangle $ are
hydrogenic states and $N$ denotes the number of microwave photons in the
occupied mode. The diagonalization of the Hamiltonian yields then
the dressed states; as its stands, the
energy spectrum is not $\Omega$-periodic.

The absorption of a circularly polarized microwave photon
increases by one the projection $L_z$
of the electron angular momentum on the $z$ axis. In the
quantum field basis this manifests itself in the fact that the
Hamiltonian matrix is block diagonal. The Hilbert space splits into
manifolds characterized by the total atom+field projection
of the angular momentum  ${\cal N}$, i.e., basis states coupled within a given
block have fixed ${\cal N}=m+N$.
${\cal N}$ is simply
the total angular momentum projection on the $z$ axis carried
by both the electron and the photons.

When the number of photons is very large, $N\gg 1,$ we can neglect
the variations in the non-diagonal matrix elements -- proportional to
either $\sqrt{N}$ or $\sqrt{N+1}$ -- across neighboring $N$ blocks.
Then, $\sqrt{2}V\sqrt{N}$ can be identified as the amplitude of the microwave
field $F.$ With this approximation, the energy spectrum is (locally)
$\Omega$-periodic. As a matter of fact, the matrix elements of $H_{\rm q}$ are
obviously equal to the ones of the classical Floquet Hamiltonian,
Eq.~(\ref{hamfloq}), already discussed in section \ref{floquet},
provided that the quantum number $K$ labelling the Fourier components in
the Floquet picture is identified with the number of photons $N.$
This establishes -- in the limit of a large number of photons --
the one-to-one correspondence between the quasi-energy
spectrum of the Floquet Hamiltonian and the energy spectrum of the dressed
atom, and also the correspondence between the Floquet eigenstates and the dressed
atom eigenstates. The two points of view are completely equivalent:
in the following, we will use the dressed atom picture as it seems
more convenient for incorporating the spontaneous emission of
photons.

As a final remark, let us note that the previous discussion is already
well known in a different context in quantum optics: for the simple case
of a two level atom coupled resonantly by an oscillatory
field, in the rotating wave approximation. There, a field-quantized
or Floquet approach yields an energy spectrum periodic in $\Omega$ 
-- a sequence
of doublets separated by the Rabi frequency \cite{cct} -- while
semiclassically in the ``rotating frame" \cite{rotframe}
one gets a single doublet of
states.

\subsection{Numerical calculation of the energy spectrum}

As discussed in the three preceding sections, there exists a
one-to-one correspondence between the eigenstates in the rotating
frame, the Floquet eigenstates and the dressed states in the quantum language. 
Therefore, we
shall call in the following the eigenenergies of either of the Hamiltonians
(\ref{hamfloq}),(\ref{hrot}) or (\ref{hq}) as quasienergies while referring 
to the eigenstates
as Floquet eigenstates or dressed states.

As we are interested in the regime of highly excited Rydberg states
(with typical principal quantum number 60) driven by a resonant microwave
field, the number of atomic states significantly coupled by the
microwave field is enormous. For computational efficiency, it is better
to use a Sturmian basis \cite{dzb95,zdb95,zdb98} rather than the
usual atomic basis. The advantage is that
it provides a discretization of the atomic continuum. Combined
with the complex rotation technique, it allows us to find not only the energies of
the dressed states but also their lifetimes
against ionization by the microwave field \cite{zdb95,zdb98}.
To numerically find several converged dressed states, 
diagonalizations of very large
matrices (of the order of $10^5$) are necessary.
As discussed above, there exists a full equivalence between
diagonalization of the Floquet Hamiltonian, i.e. clever use of the
time-periodic structure of the problem and diagonalization of the
Hamiltonian in the rotating frame.
The computational efficiencies
of the two points of view are strictly equal. Our computer code 
\cite{dzb95,zdb95,zdb98}
diagonalizes in fact $H_{\rm rot},$ contrary to the 
statement in \cite{lbf}.

\subsection{Non-dispersive wave packets}

Some time ago, Klar \cite{klar} noticed that the Hamiltonian (\ref{hrot}) 
in the rotating frame allows for
the existence of a stable fixed (equilibrium) 
point in a certain range of microwave 
amplitude $F$.
The beautiful contribution of \cite{bke} was to notice that wave
packets initially
localized in the vicinity of this fixed point will not disperse
(being bound by the fact that the fixed point is stable) for very long
time.
In the laboratory frame, these wave packets also
called ``Trojan states" in \cite{bke} 
appear as wave packets moving around the
nucleus along the circular trajectory [the periodic orbit of (\ref{hcl})].
The existence of ``stationary" wave packets -- keeping exactly their
shape during propagation along the circular orbit and hence not dispersing
at arbitrarily long times (neglecting the very slow ionization) -- was
later found in \cite{dzb95}. These are simply some among
the eigenstates of the dressed atom: by construction, they are strictly
periodic functions of time.

Semiclassically, the wave packets are manifestations of the motion inside
the stable primary 1:1 resonance zone when the microwave frequency matches
the electronic frequency. Similar states have been studied earlier for model
one-dimensional driven systems \cite{holthaus}. While for other models,
the wave packet resembles itself every period of the driving perturbation
(while changing shape within one period \cite{ab2,holthaus}), the wave
packets in circular polarization keep the same shape at all times. This is
due to the conservation of  ${\cal N}$. 

Quantum mechanically, any dressed eigenstate may be viewed as a linear
combination of atomic states, combined with appropriate photonic states
(in such a way that in the combination ${\cal N}$ is conserved). 
The wave packet
states are linear combinations of atomic Rydberg states coupled by resonant
microwaves. Thus, the biggest contribution to this combination comes from
circular states with principal quantum number $n$ close to:
\begin{equation}
n_0=\Omega^{-1/3}.
\label{n0}
\end{equation}
Note that $n_0$ is not necessarily an integer.
For the
particular case of the ``ground state wave packet" (with maximum
localization near the fixed point), it is a superposition of mainly
circular atomic states $|n,n-1,n-1\rangle$ with a small contamination 
of other
quasi-two-dimensional states of the form $|n,l,l\rangle$ and even smaller
contributions from states $l\ne m$ extending significantly out of the
$xy$ plane.

The fixed point in the rotating frame is located at $y_e=z_e=0$ and
$x_e$ satisfying
\begin{equation}
1/x_e^2+F-x_e\Omega^2=0.
\end{equation}
It is convenient
to introduce \cite{bke} a dimensionless parameter:
\begin{equation}
q=\Omega^{-2}x_e^{-3}.
\label{q}
\end{equation}
Then $q=1$ for $F=0$ and the fixed point is stable for $q\in [8/9,1]$.
It is also useful to define scaled variables which make the discussion of
the results less dependent on the particular microwave frequency value
\cite{koch}. 
We have already defined the effective principal quantum number 
$n_0$, Eq.~(\ref{n0}).
The scaled microwave amplitude is defined as 
\begin{equation}
F_0=F
n_0^4=F\Omega^{-4/3}. 
\label{scaled}
\end{equation}
$F_0$ is a dimensionless
parameter simply
related to $q$, see Eq.~(\ref{q}), via
\begin{equation}
F_0=(1-q)q^{-1/3}.
\label{scaledq} 
\end{equation}

A possible way for creation of a non-dispersive wave packet consists of the 
preparation of a circular state
with a given principal quantum number $n_0$ followed by a smooth turn on of
the microwave with frequency $\Omega=1/n_0^3$ \cite{dzb95}. Such an
approach has been tested numerically by integration of the time-dependent
Schr\"{o}dinger equation \cite{zd97}. It has been shown that the wave packet
may be obtained with about 90\% efficiency.

During the switch-on of the microwave, two regimes may be identified. For
$q$ only slightly less than unity, (small microwave amplitude), the
resonant coupling between circular states is important only. The
effective approximate Hamiltonian in the vicinity of the fixed point
corresponds then to the Hamiltonian of a pendulum \cite{ke}. This is the
standard description of the resonance at first order \cite{lili}.
For low $F_0,$ the
resonance island is small in action variable and, in effect, the
pendulum states must be quite extended in the conjugate angle. 
In the rotating frame, this manifests itself as the fixed point
being marginally stable along
the $y$ direction at vanishing $F_0$. For too small $F_0$,
the binding force is too small and the wave packet eigenstate 
can be hardly considered to be localized in the very vicinity of
the fixed point and, in turn, cannot be considered as a localized
wave packet in the laboratory frame.

With increasing microwave amplitude, the resonance island grows and
may support a few (depending on the size of the effective $\hbar=1/n_0$)
states. Then, for a sufficiently large $F_0$ (but not large
enough to destabilize the resonance), the low excitations in the
resonance island
may be well described in the harmonic approximation. It is this regime
of $F_0$ values which is of interest. 

The harmonic expansion of the Hamiltonian $H_{\rm rot}$ in rotating
frame around the equilibrium point allows one to find
normal modes of the problem and
obtain analytic predictions concerning the energies of wave packet eigenstates
\cite{bke,dzb95}.

\subsection{Harmonic approximation}

The Hamiltonian $H_{\rm rot}$ 
can be expanded
in powers of position and momentum operators in the vicinity
of the fixed point. The first order  terms cancel because
this is a fixed point. In the harmonic approximation,
only second order terms are kept (including of course crossed
position-momentum terms) and we obtain:
\begin{eqnarray}
H_{\rm rot} \!\approx\! H_{\rm h}&&=E_e +
\frac{{\bf p}^2}{2} + \Omega(\tilde{x}p_y-\tilde{y}p_x)
+ \frac{\Omega^2 q \tilde{y}^2}{2}
\nonumber\\ && - \Omega^2 q \tilde{x}^2
+ \frac{\Omega^2 q \tilde{z}^2}{2}\!  
\end{eqnarray}
where 
\begin{equation}
E_e = \frac{1-4q}{2q^{2/3}}\Omega^{2/3}
\end{equation}
is the energy of the equilibrium point and $(\tilde{x},\tilde{y},
\tilde{z})=(x-x_e,y-y_e,z-z_e)$ denotes the displacement with respect
to the fixed point.

The ``diagonalization" of this Hamiltonian leads to the
definition of the normal modes for the electronic motion. 
The calculation is tedious but straightforward:
the frequencies of the 3 normal modes labelled by the
$(+,-,z)$ indices are:
\begin{eqnarray}
 \omega_\pm &=& \Omega \sqrt{\frac{2-q\pm Q}{2}},\label{omegas}\\
\omega_z &=& \Omega \sqrt{q}
\label{ompmz}
\end{eqnarray}
with a shorthand notation
\begin{equation}
Q=\sqrt{9q^2-8q}
\end{equation}
The corresponding creation and annihilation operators in the 3 normal 
modes are linear combinations of position and momentum operators. In the harmonic
approximation,
the motion perpendicular to the plane of microwave polarization decouples
from the motion in the plane; hence, the annihilation operator in the $z$-mode
has the following simple expression:
\begin{equation}
b_z=\frac{q^{1/4} \sqrt{\Omega} \tilde{z} + i q^{1/4} p_z/\sqrt{\Omega}}
{\sqrt{2}}
\label{bz}
\end{equation}
The creation operator $b_z^{\dagger},$ being the adjoint of $b_z,$ has a
similar expression with imaginary part of opposite sign. 
The $\pm$ modes completely entangle the various coordinates and momenta
$(\tilde{x},\tilde{y},p_x,p_y).$ 
They are neither the usual ``cartesian" normal modes
along $\tilde{x}$ and $\tilde{y}$, nor ``circular" normal modes 
along $\tilde{x}\pm i\tilde{y}$, but
intermediate combinations. The annihilation operators are:
\begin{eqnarray}
\displaystyle b_{\pm}=&N_{\pm} \left( \sqrt{\Omega} \tilde{x} 
- i \frac{(Q\pm 3q\mp 2) \Omega^{3/2}}{2\omega_{\pm}} \tilde{y} 
\right.\nonumber\\
\displaystyle &\left.- i \frac{(Q\pm q)\sqrt{\Omega}}{2q\omega_{\pm}} \tilde{p_x} 
+ \frac{3q\pm Q}{2q\sqrt{\Omega}} \tilde{p_y}
\right),
\label{bpm}
\end{eqnarray}
with the normalization constants given by:
\begin{equation}
N_{\pm} = \left( \frac{q^2(2Q^2-q+2 \pm 3(1-2q)Q)}
{8Q^2(1-q)} \right)^{1/4}.
\end{equation}
The Hamiltonian in the harmonic approximation 
can be expressed
in terms of creation and annihilation operators of the electronic normal modes 
\cite{bke,dzb95,zdb95}:
\begin{eqnarray}
H_{\rm h}
& =& E_e \! + \! \omega_+ (b^\dagger_+ b_+ +1/2)\nonumber\\
&&- \omega_- (b^\dagger_- b_-+1/2)
 + \omega_z (b^\dagger_z b_z+1/2).
\label{hh}
\end{eqnarray}

Note the unusual minus sign in front of the $\omega_-$ term. This comes
from the fact that the stable fixed point is not a minimum of the
Hamiltonian. However, the presence of crossed momentum-position terms
assures the complete stability of the equilibrium point versus
any small perturbation, in complete analogy with a usual
potential minimum.

Let us denote the atomic eigenstates in the harmonic approximation as
$|n_+,n_-,n_z)$ where $n_i$ denotes the excitation in a given $\pm,z$ mode
\cite{dzb95,zdb95}. We will also -- when needed -- denote the Floquet
or dressed atom
eigenstates (thus including the microwave field degree of freedom)
as $|n_+,n_-,n_z;{\cal N})$. The corresponding energies are:
\begin{eqnarray}
E(n_+,n_-,n_z;{\cal N}) &=& E_e + {\cal N} \Omega 
+ \left(n_++\frac{1}{2}\right)\omega_+
\nonumber\\
&& - \left(n_-+\frac{1}{2}\right)\omega_-
+  \left(n_z+\frac{1}{2}\right)\omega_z .
\label{enharm}
\end{eqnarray}

It has been verified that exact Floquet quasienergies,
corresponding to dressed states localized in the vicinity of the
equilibrium point in the rotating frame agree with the harmonic prediction,
Eq.~(\ref{enharm}),
to within 10\% of the mean level spacing \cite{dzb95,zdb95} for wave
packets corresponding to $n_0\approx 60$ or higher.
This indicates that the harmonic approximation is quite satisfactory
for sufficiently large $n_0$ (sufficiently small $\Omega$).
In particular,
the ``ground state'', $|0,0,0)$ corresponds to the most
tightly bound state, centered at the fixed point in the rotating frame.
Its wavefunction
has a
Gaussian shape in the harmonic approximation. The corresponding
exact Floquet (dressed) state \cite{zdb95}
 is a non-dispersive wave packet 
since in the laboratory frame it
moves in time around the nucleus with frequency $\Omega$ following
the stable circular periodic orbit. It is precisely the spontaneous
emission from this state which will be discussed in the following sections.

The
partial scheme of levels resulting from $H_h$ is shown in Fig.~\ref{levels}
for a few  values of the ``total'' angular momentum ${\cal N}$.
The neighboring ladders of states are shifted with respect to each
other by the microwave frequency $\Omega$ corresponding to different
decoupled blocks of the Floquet Hamiltonian. Transitions
between quasienergy levels
(Floquet or dressed states) may be introduced by some additional
perturbation. In particular one may consider the so-called Floquet
spectroscopy \cite{munchen}, absorption or stimulated emission between
dressed states introduced by an additional laser or microwave field. 

 It seems quite
natural to study the possible effects of spontaneous emission of these
states within the very same harmonic approximation. This allows to obtain
close analytic expressions for the spontaneous decay rate.
The validity of the harmonic
approximation has been challenged strongly by Farrelly and coworkers
\cite{lbfl,flu,buf,lbf,clfu}. By comparison of spontaneous spectra obtained
in the harmonic approximation with full quantum mechanical results, we
show later that their claims are not justified.

\section{Spontaneous emission of the wave packet in the dressed atom picture}

\subsection{General properties}

\label{spont}

To consider the spontaneous emission one has to consider the atom
interacting not only with a single mode [treated either classically
(\ref{hcl}) or quantum mechanically (\ref{hq})] but a full electromagnetic
field. The corresponding Hamiltonian reads, in atomic
units~\cite{foot6}
\begin{equation}
H=H_{\rm i}+{\bf r}\cdot{\bf E}({\bf 0}) +H_{\rm f}
\label{hsp}
\end{equation}
where subscript ${\rm i}$ stands for ${\rm cl}$ or ${\rm q}$ 
corresponding to
the classical or quantum version of the description of the
occupied microwave mode, respectively. ${\bf r}$ is the usual dipole
operator,
$\bf{E(r)}$ denotes the electric field operator at position $\bf{r}$ and 
$H_{\rm f}$ is the
free field Hamiltonian. Both $H_{\rm f}$ and $\bf{E}$ can be expanded
over modes of the field. While in \cite{bb97}, the expansion
in sperical waves
 is used (which facilitates explicit transformation to the
rotating frame), we expand the field into standard plane waves~\cite{cct}.
We normalize the modes per unit in ${\bf k}$ space (related to the
frequency in atomic units by $\omega=\alpha k$ where 
$\alpha=e^2/\hbar c\simeq
1/137.036$ is the fine structure constant), then
\begin{equation}
{\bf E}({\bf r})=\sum_j{\int{\frac{\sqrt{\omega}}{2\pi}
\left(a_j({\bf k}){\bf e}_j\exp(i{\bf k}\cdot{\bf r}) +{\rm h.c.}\right)
\ d^3{\bf k}}},
\label{field}
\end{equation}
 where h.c. stands for hermitian conjugation, $a_j({\bf k})$ 
 is the annihilation
 operator in the mode characterized by wavevector ${\bf k}$ and
 polarization ${\bf e}_j$ (the sum over $j$ is over the two
 possible orthogonal polarizations). Note that in (\ref{hsp}) the electric
 field operator is evaluated at origin, consistently with the dipole
 approximation assumed in this paper. Similarly the free field 
 hamiltonian is
 \begin{equation}
 H_{\rm f} = \sum_j{\int{\omega 
 a_j({\bf k})^\dagger a_j({\bf k})\ d^3{\bf k}}}.
 \end{equation}
 
Now we can use the Fermi golden rule to calculate
the decay rate due to spontaneous emission
from an arbitrary initial state $|i\rangle$ of the atom
dressed by the microwave field to any final state $|f\rangle$,
taking into
account the density of modes
for the photons \cite{foot4}.
The result is:
\begin{equation}
\Gamma = \frac{4\alpha^3 \omega^3}{3} |(f|{\bf r}|i)|^2
\label{gamma}
\end{equation}
where $\omega$ is the energy difference between the initial and final
dressed states.

\begin{figure}
\centerline{\psfig{figure=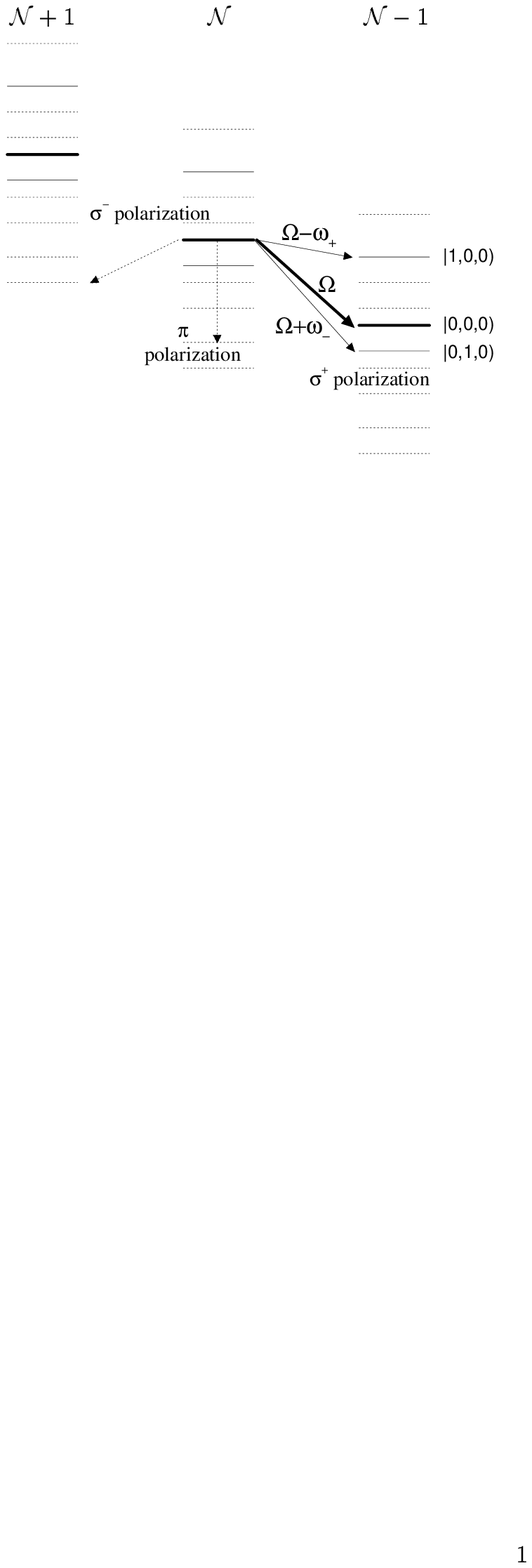,width=8cm,bbllx=100pt,bblly=530pt,bburx=300pt,bbury=770pt}}
\medskip
\caption{Schematic energy levels of an hydrogen atom dressed by
a microwave field of circular polarization, with Hamiltonian
given by Eq.~(\protect{\ref{hq}}). The total angular momentum
${\cal N}$ 
(atom+field) is a constant of motion, so that the spectrum can be split
in separate ladders. The consecutive ladders have the same structure,
simply separated (in energy) by the microwave frequency $\Omega.$
We have here represented three consecutive ladders. 
If all
ladders are taken into account, the spectrum is $\Omega$-periodic
and coincides with the Floquet spectrum of the classical Hamiltonian,
Eq.~(\protect{\ref{hamfloq}}). Alternatively, the 
spectrum in each ladder can be thought as the energy spectrum of the
Hamiltonian in rotating frame, Eq.(\protect{\ref{hrot}}).
Spontaneous emission of photons
(represented by arrows in the figure)
 from a state in ladder ${\cal N}$
is only possible to another state lying at
a {\em lower energy} in either ladder ${\cal N}-1$ (with 
$\sigma^+$ polarization), ladder ${\cal N}$ (with $\pi$ polarization)
or ladder ${\cal N}+1$ (with $\sigma_-$ polarization).
 Hence, $\sigma^+$ transitions are favored.
In the specific case of the ground non-dispersive wave packet
$|0,0,0)$ in the harmonic approximation, only three
transitions, represented by the solid arrows, are allowed at frequencies
$\Omega-\omega_+$, $\Omega$ and $\Omega+\omega_-$, all of them
being $\sigma^+$ polarized.}
\label{levels}
\end{figure}

Spontaneous emission will take place
between the various states belonging to ladders depicted in
Fig.~\ref{levels}. In the process, the projection of the full
angular momentum on the $Oz$ axis is conserved. Here, this total
angular momentum has three contributions: the atomic angular
momentum $m$, the angular momentum $N$ of the driving microwave
photons (all in the same mode) and the angular momentum $M$ in the
other modes of the electromagnetic field. Hence, we have
${\cal M}={\cal N}+M=N+m+M=$ constant. As spontaneous emission is here
considered as a weak perturbation, it will simply induce transitions
between eigenstates of the dressed atom, see Eq.~(\ref{hq}), for which
${\cal N}$ is a good quantum number (see preceding section).
In the dipole approximation, for a one-photon process, the
atomic angular momentum $m$ can change by at most one. 
When the spontaneous photon is emitted in an initially empty mode
(which is the process we are interested in), the number of photons
in the driving microwave mode $N$ does not change. Hence, ${\cal N}$
changes by at most one, and the same is true (with opposite sign) for
$M$. 
 
Thus, in the dipole approximation,
there are only three possible processes. If the spontaneous
photon is emitted with
$\sigma^+$ polarization, i.e. angular momentum $M=1$,
it will induce a transition
from a state in a given ${\cal N}$ ladder to a state in the 
${\cal N}'={\cal N}-1$ ladder. Similarly, for $\sigma^-$ polarization, 
i.e. angular momentum $M=-1$,
it will induce a transition from ${\cal N}$ to ${\cal N}'={\cal N}+1$ and
for $\pi$
polarization, $M=0$ and ${\cal N}'={\cal N}$. 
As all but one photon modes are initially empty, by spontaneous emission,
the energy $\hbar\omega$ is gained in one mode and thus has to be lost
by the dressed atom.
Note that here and in the rest of this paper, we discuss the polarization
of the various transitions using a quantization axis along the $z$ axis,
the
direction of the propagation of the driving field.
 As explained below the spontaneous photons
are mainly $\sigma^+$ polarized with respect to the $z$ axis. Hence,
the helicity of the photons (with respect to their direction of emission, i.e.
the quantity measured in an experiment) is in
general elliptical, depending on the direction of emission.

In summary, in the dressed atom picture as shown in Fig.~\ref{levels},
the only possibilities for emitting a spontaneous photon is to
decay to a state with lower energy in the same ladder or in one
of the two neighboring ladders. All other processes are forbidden.
Note however that -- in the absence of a further approximation --
plenty of processes with all 3 possible polarizations should exist.
However, as can be seen from Fig.~\ref{levels}, 
the $\sigma^+$ emission is
clearly favored, since the ${\cal N}-1$ ladder is shifted downwards
with respect to the initial state by $\Omega.$ $\pi$ is less favored
and $\sigma^-$ even less because of the upwards energy shift.

\subsection{Spontaneous emission from a non-dispersive wave packet
in the harmonic approximation}

The general properties discussed in the preceding section are valid
whatever initial state is chosen. When the initial state is a
non-dispersive wave packet and when the scaled microwave field
$F_0$ is sufficiently large, we can use the harmonic approximation
described above. 
We will make this approximation throughout the rest of this section.
The atomic dipole components $x\mp iy,z$ corresponding respectively
to the photon polarizations $\sigma_{\pm},\pi$ can be expressed as functions
of creation/annihilation operators in the normal modes, 
Eqs.~(\ref{bz})-(\ref{bpm}):
this is straightforward as all these operators are linear combinations of
positions and momenta.
The result is:
\begin{eqnarray}
x\pm iy & = &  \Omega^{-2/3}q^{-1/3}\nonumber\\
&& + \frac{1}{4\sqrt{\Omega}Q}
\left( 
\frac{-3q+Q\pm (q-Q)\frac{\omega_+}{\Omega(1-q)}}
{N_+} b_+^{\dagger}\right.\nonumber \\
&&+ \frac{-3q+Q\mp (q-Q)\frac{\omega_+}{\Omega(1-q)}}
{N_+} b_+\\
&&+ \frac{3q+Q\pm (q+Q)\frac{\omega_-}{\Omega(1-q)}}
{N_-} b_-^{\dagger}\nonumber \\
&&\left.+ \frac{3q+Q\mp (q+Q)\frac{\omega_-}{\Omega(1-q)}}
{N_-} b_- \right)\nonumber\\
z & = & \frac{q^{-1/4}}{\sqrt{2\Omega}}(b_z^{\dagger}+b_z)
\label{expansion}
\end{eqnarray}

Hence, in the harmonic approximation, there are only very few transitions
allowed. The selection rules are:
\begin{equation}
\begin{array}{lll}
\Delta n_+=0\ \ &\Delta n_-=0,\pm1\ \ &\Delta n_z=0\ \ {\rm or}\\
\Delta n_+=0,\pm 1\ \ &\Delta n_-=0\ \ &\Delta n_z=0
\end{array}
\label{sr1}
\end{equation}
in $\sigma^+$ and $\sigma_-$ polarizations and:
\begin{equation}
\Delta n_+=0\ \ \Delta n_-=0\ \ \Delta n_z=\pm1
\label{sr2}
\end{equation}
in $\pi$ polarization.

This means that a given $|n_+,n_-,n_z;{\cal N})$ dressed eigenstate
is connected to at most 2 states in the same ${\cal N}$ ladder and
5 states in each of the ${\cal N}\pm1$ neighboring ladders.
Among these 12 possible transitions, only those ending in a final
state with lower energy are allowed in a spontaneous emission process.
Using Eq.~(\ref{enharm}), we see that the energy loss is
$\pm\omega_z$ for the 2 $\pi$-polarized transitions,
$\Omega,\Omega\pm\omega_+,\Omega\pm\omega_-$ for the 5 $\sigma_+$
transitions and $-\Omega,-\Omega\pm\omega_+,-\Omega\pm\omega_-$ 
for the 5 $\sigma_-$ transitions. From 
Eq.~(\ref{ompmz}), we know that $\omega_+$ and $\omega_-$ are both
smaller than $\Omega.$ Hence, all 5 transitions in $\sigma_-$ polarization
are forbidden (by energy conservation) for spontaneous emission.
On the other hand, all 5 transitions in $\sigma_+$ polarization
are allowed. In $\pi$ polarization, only the transition $\Delta n_z=-1$
is allowed by energy conservation.
To summarize, in the harmonic approximation, a given  $|n_+,n_-,n_z)$
atomic eigenstate can only emit spontaneous photons at the six
frequencies $\omega_z,\Omega,\Omega\pm\omega_+,\Omega\pm\omega_-$
the first transition being $\pi$ polarized, the 5 other ones having
$\sigma_+$ polarization.

In the specific case where the initial state in the ground
state wave packet $|0,0,0)$ with minimum values of the
$n_+,n_-,n_z$ quantum numbers  and maximum localization around
the equilibrium point, all transitions {\em decreasing} any
of the $n_+,n_-,n_z$ are obviously non-existent.
Hence, we are left with only 3 possible transitions, all of them
being $\sigma_+$ polarized:

\begin{itemize}
\item Transition from the ground state wave packet $|0,0,0;{\cal N})$ to its
image $|0,0,0;{\cal N}-1)$ at microwave frequency $\Omega$.
While the atomic state is not changed, a photon of the driving microwave
is elastically scattered into another mode at the same frequency, but
with a different direction. In the language of quantum optics,
this process can be viewed as the elastic component in the 
resonance fluorescence of the driven atom.
\item Transition from $|0,0,0;{\cal N})$ to a $n_+=1$ dressed state
$|1,0,0;{\cal N}-1)$ occurring at frequency $\Omega-\omega_+$. This is
an inelastic component.
\item Transition from $|0,0,0;{\cal N})$ to a $n_-=1$ dressed state
$|0,1,0;{\cal N}-1)$ occurring at frequency $\Omega+\omega_-$. This is also
an inelastic component. Note that the difference on signs
in the frequencies of the two inelastic components, $\Omega-\omega_+$ and
 $\Omega+\omega_-$ is directly related to the unusual - sign in the
Hamiltonian, Eq.~(\ref{hh}).
\end{itemize}

In the harmonic approximation, we can go beyond the above
qualitative predictions and calculate the partial decay rates
to the different final states analytically. To this end, we need
the matrix elements of the dipole operators between the initial
and final states. This is easily obtained using the expansion of the
dipole operator on the creation/annihilation operators in the 
normal modes, Eq.~(\ref{expansion}), as the
matrix elements of these operators between eigenstates are well known.
We obtain:
\begin{eqnarray}
(0,0,0|x-iy|0,0,0) & = & \Omega^{-2/3}q^{-1/3} \\
(1,0,0|x-iy|0,0,0) & = & 
\frac{-3q+Q-(q-Q)\frac{\omega_+}{\Omega(1-q)}}
{4\sqrt{\Omega}QN_+}\\
(0,1,0|x-iy|0,0,0) & = &
\frac{3q+Q-(q+Q)\frac{\omega_-}{\Omega(1-q)}}
{4\sqrt{\Omega}QN_-}.
\label{dipole}
\end{eqnarray}

This finally makes it possible to calculate the decay rates of the
3 spontaneous emission processes, using Eq.~(\ref{gamma}):
\begin{equation}
\Gamma(\Omega) = \frac{2 \alpha^3 \Omega^{5/3} q^{-2/3}}{3}
\label{decayel}
\end{equation}
for the elastic component,
\begin{equation}
\Gamma(\Omega-\omega_+) = \frac{2 \alpha^3 (\Omega-\omega_+)^3}{3}
|(1,0,0|x-iy|0,0,0)|^2
\label{decayp}
\end{equation}
and
\begin{equation}
\Gamma(\Omega+\omega_-) = \frac{2 \alpha^3 (\Omega+\omega_-)^3}{3}
|(0,1,0|x-iy|0,0,0)|^2
\label{decaym}
\end{equation}
for the two inelastic components.
The various quantities in these formula are related to the
microwave frequency $\Omega$ and to the scaled parameter $q$
(itself related to the microwave amplitude by Eqs.~(\ref{scaled}) and
(\ref{scaledq})) by Eqs.~(\ref{ompmz}) and (\ref{dipole}).
For a fixed classical dynamics (fixed $q$ i.e. fixed scaled
microwave amplitude $F_0$), the elastic decay rate
scales as $\Omega^{5/3}=n_0^{-5}$ as does the decay
rate of a circular state while the inelastic decay rates
scale differently as $\Omega^{2}=n_0^{-6}.$
The results can be converted to usual units by multiplying
them by
$me^4/\hbar^3\simeq 4.13 \times 10^{16}\ {\rm s}^{-1}.$

\begin{figure}
\centerline{\psfig{figure=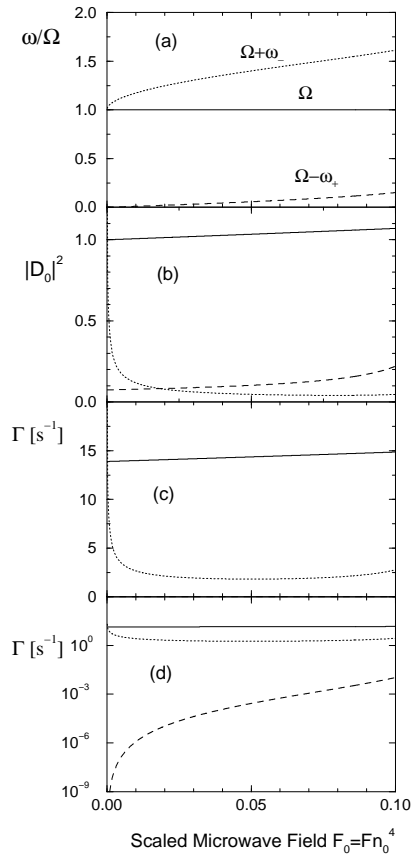,width=9cm}}
\medskip
\caption{Analytic predictions for the spectrum of spontaneous photons
emitted by a non-dispersive wave packet $|0,0,0)$ of the hydrogen atom in 
a cicularly polarized microwave of frequency
$\Omega/2\pi=30.46\ {\rm GHz}$ corresponding to a principal
quantum number $n_0=60$, as a function of the scaled microwave field
amplitude.
Solid line corresponds to elastic scattering of the microwave photons,
the dashed and dotted lines correspond to inelastic processes.
 (a): Frequencies of the emitted photons.  (b): Squares of the
dipole matrix elements involved in the 3 transitions. The elastic
component is largely dominant. (c): Decay rates along the 3 transitions.
Because of the rapid increase in the density of electromagnetic modes
with frequency, the $\Omega-\omega_+$ component is almost invisible.
(d): Same as (c) on a logarithmic scale.
The decay rates of the $\Omega-\omega_+$, $\Omega$ and $\Omega+\omega_-$
components roughly scale  as $F_0^3n_0^{-6}$, $n_0^{-5}$
and $F_0^{-1/2}n_0^{-6}$, respectively.}
\label{fdep}
\end{figure}

In Fig.~\ref{fdep}, we show the frequencies, squared dipole matrix
elements and decay rates (in linear and logarithmic scales) of the three
components as a function of the scaled microwave field for principal
quantum number $n_0=60,$ the corresponding microwave frequency being
$\Omega/2\pi=30.46\ {\rm GHz}.$
The first thing to note is obviously the dominance of the elastic
component at the microwave frequency. About 80\% of the spontaneous
photons emitted by the system are at the microwave frequency. Elastic
scattering of the microwave photons to other modes (different direction)
is by far the dominant process. The radiation diagram of these photons has
the standard dipolar shape (this is identical for the three components as they
are all due to the $x-iy$ dipole operator).
This dominance is easily understood from the different
scalings of the three decay rates, 
Eqs.~(\ref{decayel})-(\ref{decaym}) discussed above.
The dipole matrix element $(0,0,0|x-iy|0,0,0)$ measures the static dipole,
i.e. the position of the center of the non-dispersive wave
packet. It is thus -- in the harmonic approximation - equal to 
$x_e$, the coordinate of the classical fixed point, of the
order of $n_0^2.$ In contrast, the non diagonal matrix elements
$(0,1,0|x-iy|0,0,0)$ or $(1,0,0|x-iy|0,0,0)$ measure how the excited
states explore the vicinity of the fixed point. They are of the order
of $n_0^{3/2},$ i.e. about $\sqrt{n_0}$ times smaller. This is why
the corresponding decay rates are much smaller. Because of the different
scaling properties of the decay rates, it is easy to check that a change
in the value of $n_0$ gives similar plots for the decay rates of the
inelastic components as a function of the scaled microwave field
(they are just multiplied by a uniform numerical factor) while
the elastic component is relatively increased in proportion with $n_0.$
In other words, in the semiclassical limit $n_0\rightarrow \infty,$
the inelastic components asymptotically vanish.

Another important observation is that, although the two inelastic
components have dipole matrix elements of comparable magnitude
(see Fig.~\ref{fdep}b), the decay rate of the $\Omega-\omega_+$
component is several orders of magnitude smaller than the decay rate
of the $\Omega+\omega_-$ component, see Fig.~\ref{fdep}c. 
This is entirely due to the
$\omega^3$ factor representing the density of electromagnetic mode.
In practice, the $\Omega-\omega_+$ component
will be very difficult to detect.

Finally, when the microwave field is sufficiently small,
the various quantities can be expanded in powers of the scaled field.
In this limit (see also Fig.~\ref{fdep}d), the $\Omega-\omega_+$ decay
rate goes to zero as $F_0^3$ whereas the $\Omega+\omega_-$
decay rate increases like $1/\sqrt{F_0}$ and the elastic
component $\Omega$ keeps a constant decay rate. Of course, such a divergence
in clearly unphysical and simply indicates the breakdown of the harmonic
approximation. At lower field, the pendulum approximation of the
Hamiltonian \cite{dzb97} has to be used, restoring finite decay rates,
roughly equally shared by the $\Omega+\omega_-$ and $\Omega$ components.

\section{Comparison with earlier results}

The former approach to spontaneous emission of the non-dispersive
wave packets \cite{bb97} has been carried out in the rotating frame
using the Hamiltonian Eq.(\ref{hrot}), or rather its generalization
which includes also other non occupied electromagnetic modes. 
Using the notion of rotational frequency shift found previously by the same
authors \cite{bb97r}, the frequency of the emission in the laboratory
frame has been found. A single frequency, since the authors considered
the emission at $\Omega+\omega_-$ only assuming, incorrectly, that
spontaneous emission of a photon in the rotating frame must lead to a
decrease of the energy (as it happens in the laboratory frame).

To see the fallacy of this argument, consider the standard spontaneous
decay of the circular $|n+1,n,n\rangle$ hydrogenic state in the absence
of any driving electromagnetic field.
In the dipole approximation, it can decay spontaneously 
only to the $|n,n-1,n-1\rangle$
circular state. Consider the same situation in the frame rotating
at frequency $\Omega$. The Hamiltonian in the rotating frame is obtained
from Eq.~(\ref{hrot}) with $F=0$. The energy of the initial state
in the rotating frame is then $E_i=-1/2(n+1)^2-\Omega n$ while
the energy of the final state $E_f=-1/2n^2-\Omega (n-1)$ is greater than
$E_i$ provided $\Omega>(n+1/2)/(n(n+1))^2\approx 1/n^3$. Hence, emission
of the spontaneous $\sigma^+$ polarized photon {\em increases} the energy
in the rotating frame.

The same situation occurs for the standard toy model of quantum optics:
the two-level atom interacting with a quasi-resonant monochromatic
field, treated in the rotating wave approximation.
In the ``rotating frame'' \cite{rotframe}, spontaneous emission
does not occur only from the upper to the lower state of the 
(Autler-Townes) doublet.
This particular line (upper to lower state in the doublet) occurs
at the frequency of the driving field plus the Rabi frequency.
But the emission may also take place from the lower to the upper state of
the doublet (at driving frequency minus the Rabi frequency) as well
as from the lower (upper) state to ``itself". In effect the
resonance fluorescence spectrum has the famous Mollow triplet structure
\cite{resonflu,cct} with the dominant elastic component in the middle.

This discussion shows that the rotating frame approach leads to
quite counter intuitive picture.
 By comparison, the dressed state
picture presented in the previous section seems to be quite
intuitive. One may easily verify that extending the calculation
presented in \cite{bb97} to include the lines emitted at
the microwave frequency $\Omega$ (which corresponds in the rotating
frame to the emission from the wave packet state to itself) as
well as at $\Omega-\omega_+$ leads to the same results as those
obtained in the dressed atom picture above \cite{foot7}. This shows that both
approaches lead, as it should be, to the same spontaneous emission
rates.

\null From a practical point of view, as the decay at $\Omega-\omega_+$
is much slower than the decay at $\Omega+\omega_-$ (see discussion above),
the numbers given in \cite{bb97} can be considered correct\cite{foot5}
if elastic scattering does not have to be taken into account.
A non-trivial question is whether elastic scattering of photons
has to considered as a decay process or not. In our opinion, this
is simply a question not properly formulated. Thinking resonance
fluorescence in terms of decay is probably inadequate as the system
is continuously fed with external photons.
If one measures the population in the ground state wave packet
$|0,0,0),$ this quantity will certainly decay because
of inelastic components only. If one measures the spectrum
radiated by the atom, the answer is completely different
as the elastic component dominates. Also, if one considers the
widths of the various components, the elastic process will come into
the game as it destroys the phase coherence of the initial wave packet
and consequently broadens the line. A general discussion of such effects
and more generally of
resonance fluorescence in the framework of quantum optics can
be found in \cite{resonflu}.

Let us also note that the ``classical'' emission rate of an
electron moving on a circular orbit presented in \cite{bb97}
is incorrect. A classical charge moving on a circular orbit
with frequency $\Omega$ must emit photons at the same frequency.
In our opinion, the only meaningful quantity is the decay rate
of the {\em energy} in the atomic system.
Classically, it is proportional to the square of the acceleration
and, for the classical periodic orbit at the microwave frequency, it is
in atomic units \cite{jackson}:
\begin{equation}
\frac{dE}{dt} = \frac{2 \alpha^3 \Omega^{8/3} q^{-2/3}}{3}
\end{equation}
Considering that, in the semiclassical limit $n_0\rightarrow \infty,$ the
system radiates essentially photons as frequency $\Omega$, we immediately
see that this classical rate {\em exactly} 
(i.e. without additional numerical factor of the order of one) coincides
with the quantum decay rate of energy deduced from Eq.~(\ref{decayel}).
It also exactly corresponds to the decay rate of atomic circular
states, as given in e.g. Ref.~\cite{bethe}, in the limit
of vanishing microwave field $q\rightarrow 1.$
In \cite{bb97}, the decay rate in the $\Omega+\omega_-$ component
is incorrectly compared to the relative decrease of the classical energy
in time. The first quantity -- as discussed above -- is roughly
$n_0$ times smaller than the elastic decay rate while the relative decrease
of classical energy in time is also about $n_0$ times smaller than
the rate of emission of photons because the energy of a radiated photon
is about $n_0$ times smaller than the total energy. The two $n_0$ dependences
cancel out, but not the numerical factors in front. This explains why
a false classical-quantum correspondence (within constant numerical factors)
is observed in \cite{bb97}. Both classical and quantum decay rates
are -- in our opinion --  underestimated by a $\simeq n_0$ factor.

\section{How good is the harmonic approximation ?}

\label{howgood}

\null From the beginning of investigations on non-dispersive wave packets
in the hydrogen atom driven by a circularly polarized microwave, 
there has been a controversy regarding the accuracy of the harmonic
approximation used when introducing the wavepackets as
Gaussians centered around the equilibrium point
 \cite{bke},
see Section II.
Firstly in a Comment \cite{flul} and later in a
number of papers \cite{lbfl,flu,buf,lbf,clfu} Farrelly
and coworkers claimed
that ``highly anharmonic nature of the effective potential in
the vicinity of the equilibrium point will cause any such putative
coherent state to disperse ..'' unless the wave packet is built from
states of unrealistically high quantum number ($n_0>200$).
They have
shown that an additional static magnetic field applied in the direction
of microwave propagation ($Oz$ axis) allows to relatively
decrease the importance of the
anharmonic terms in the expansion around the equilibrium point.
Further they claim that ``because these extrema are locally harmonic
their vacuum states are truly coherent states in the original sense of
Schr\"odinger'' \cite{clfu} (by extrema one should understand
global equilibrium points).

Our understanding of the phenomenon is slightly different. While
Farrelly and coworkers stress that the Gaussian shape of
the wave packet is essential for non spreading time evolution in the
locally harmonic potential, we adopt a bit less narrow-minded
point
of view. We have shown \cite{dzb95} that, even if the harmonic
approximation is only partially valid, there will be a one-to-one
correspondence between the quasienergy states of the approximate
harmonic Hamiltonian, Eq.~(\ref{hh}), and the full Hamiltonian in the
rotating frame, Eq.~(\ref{hrot}). In particular, the ground state
of the local harmonic approximation (referred to above as ground state
wave packet) is a good approximation to an {\it exact} 
eigenstate of the full problem (or via the unitary transformation to
the corresponding Floquet or dressed state). We refer to this
exact eigenstate of the full atom+microwave field problem as the
non-dispersive wave packet since:
\begin{itemize}
\item it is a linear combination of atomic states;
\item it is centered at the stable equilibrium point in the rotating
frame, thus it follows the classical circular periodic orbit of
the problem in the laboratory frame;
\end{itemize}
thus fulfilling standard textbook criteria for ``a wave packet''.
Being an eigenstate of $H_{\rm rot}$, Eq.~(\ref{hrot}), the state
is rigorously non spreading provided we neglect the coupling to the
continuum -- i.e. the possibility of ionization. We have shown
\cite{dzb95,zdb95,zdb98} that typical lifetimes for the
ionization exceed many thousands of Kepler periods (reaching up
to $10^6$ Kepler periods for $n_0\approx 60$ or more).

In our understanding of the phenomenon, the validity of the harmonic
approximation is not crucial for the very existence of 
non-dispersive wave packets in the Hydrogen atom driven by a circularly
polarized microwave.
Moreover, the
phenomenon is more general and, as mentioned in the introduction,
may be found in other systems such as atoms driven by a linearly
polarized microwave \cite{ab2} or one-dimensional models
\cite{holthaus} supporting stable classical 1:1 resonance islands.
In the general case, the wave packet preserves its shape at intervals
equal to the period of the external driving, only in the circular
polarization problem (due to the additional symmetry which allows for
the complete removal of the time-dependence) the wave packet shape
is the same at all times. There is no reason to restrict the
notion of non-dispersive (or non spreading) wave packet to a 
single particular situation
which can be made almost fully harmonic. One should also remember that
even in the presence
of a magnetic field, a case favored by Farrelly and coworkers,
the potential is also only approximately harmonic, and 
Gaussian wavepackets will eventually disperse. While wave packets in
our sense, being single dressed states, will not. 
In any case, the
addition of a magnetic field does not change the nature of the problem,
it may just keep some anharmonic terms smaller.

While not necessary for the very existence of non-dispersive wave packets,
the harmonic approximation is useful for analytic evaluation
of different properties, such as spontaneous emission discussed
above. In view of Farrelly and coworkers remarks, it would seem
that the results obtained in the previous section are incorrect.
The simplest way to check this statement is to evaluate the ``exact"
spontaneous
emission spectrum from the ground quasienergy eigenstate of the full
Hamiltonian.

The presence of the anharmonic terms in the full Hamiltonian has to
manifest itself in the break-up of the selection rules discussed
in Section II. Instead of three possible lines in the $\sigma^+$
emission, one should expect a multitude of lines. Using the harmonic
level assignment, the $|0,0,0;{\cal N})$ wave packet may then decay
to e.g. states approximated  
by $|i,j,k;{\cal N}-1)$ with $i,j,k$
being larger than unity (with $k$ even since $\sigma^+$ dipole element
may link states of same $z$-parity only).
 Also, due to couplings introduced by nonlinear
terms, spontaneous emission with other (i.e. $\sigma_-$ or $\pi$)
polarizations becomes possible.

\begin{figure}
\centerline{\psfig{figure=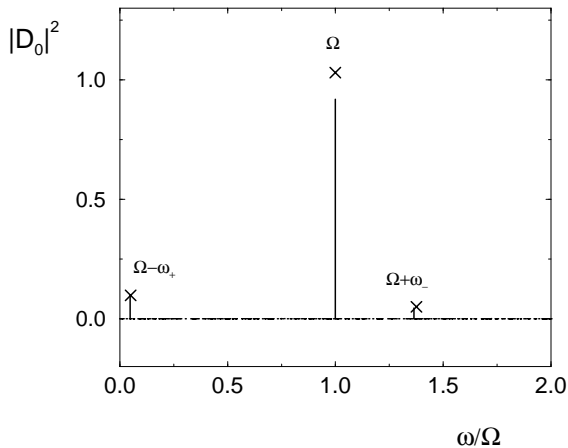,angle=-90,width=9cm}}
\medskip
\caption{Single photon spontaneous emission from the
non-dispersive wave packet of the hydrogen atom in 
a cicularly polarized microwave of frequency
$\Omega=1/60^3$ (in atomic units) corresponding to a principal
quantum number $n_0=60$, and scaled amplitude
$F_0=0.04445881$ (corresponding to $q=0.9562$). We have here represented
the square of the scaled dipole matrix element $D_0=D/n_0^2$ connecting the initial
$|0,0,0)$ wave packet to the final state as a function of the
scaled energy difference (hence, ratio of the frequency of the emitted photon
to the microwave frequency $\Omega$) between
the two states. The stick spectrum  is the ``exact" result as obtained
from a numerical diagonalization and the crosses represent  the
analytic result obtained using the harmonic approximation,
Eq.~(\protect{\ref{dipole}}). There are three dominant lines at
frequencies $\Omega-\omega_+$, $\Omega$ and $\Omega+\omega_-$, all of them
$\sigma^+$ polarized. The strongest line (by a factor proportional to
$n_0$) is the elastic scattering of the microwave photons by the
dressed atom. The other transitions (as well as transitions with
$\sigma_-$ or $\pi$ polarizations) are negligible, the corresponding sticks
have practically zero heights (yielding a broken line build of individual sticks
each representing one of the over 400 Floquet final states), which proves
the validity of the harmonic approximation. 
}
\label{exdip}
\end{figure}

Thus, the spontaneous emission spectrum provides a useful test of
the validity of the harmonic approximation. Having at our disposal
the computer code which allows reliably to find the exact quasienergy
states of the problem \cite{dzb95,zdb95,zdb98}, it has been sufficient
to evaluate dipole matrix
elements between the exact quasienergy states.
The resulting squared dipole moments are shown in Fig.~\ref{exdip} together
with the corresponding squared dipoles as given by the harmonic approximation,
Eq.~(\ref{dipole}),
presented as three crosses.
 Observe that,
the harmonic approximation works quite well. The
selection rules are well preserved, the frequencies of the emission
(line positions) agree very well, the dipole values (heights of
the lines) agree within 10\% for all three lines.
This shows that the harmonic
approximation is quite good, {\it in the absence of any magnetic
field,} not only for the ground state wave packet, $|0,0,0)$
but also for primary $\omega_+$ [state $|1,0,0)$] and $\omega_-$
[state $|0,1,0)$] excitations. Observe that the height of the
peak at $\Omega$ is proportional to $|(0,0,0|x-iy|0,0,0)|^2$ involving
the wavefunction of the ground state wave packet only. The sidebands,
corresponding to transitions to excited wave packets provide
independent tests of the corresponding wavefunctions.

\begin{figure}
\centerline{\psfig{figure=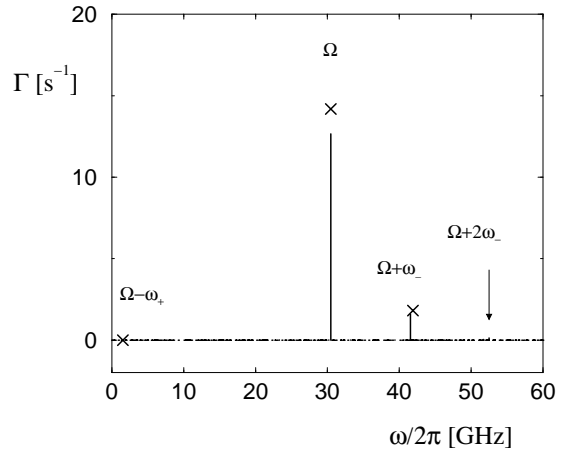,angle=-90,width=9cm}}
\medskip
\caption{
Same as figure \protect{\ref{exdip}}, but for the
decay rates along the different transitions and in S.I. units. 
The density of modes
of the electromagnetic wave introduces an additional $\omega^3$ 
factor which completely kills the transition at frequency $\Omega-\omega_+$,
invisible in the figure. One can see a  small line at frequency approximately
$\Omega+2\omega_-$ (see arrow), an indication of a weak breakdown of the harmonic
approximation.
}
\label{exsp}
\end{figure}

Fig.~\ref{exsp} shows the corresponding spontaneous emission rates.
The component at $\Omega-\omega_+$ is not visible due
to the low frequency of the emission. On the other hand, the spectrum
at high frequencies is enhanced by cubic power of the frequency
factor
coming from the photonic density of states. Here, careful inspection
of the spectrum corresponding to exact eigenstates reveals a tiny
peak centered at $\Omega+2\omega_-$ (indicated by an arrow in the figure). 
Such a transition is forbidden
in the harmonic approximation, its presence indicates that the harmonic
approximation is not exact. Still, the smallness of anharmonic terms
strongly indicates that the harmonic approximation works
well, at least for the typical value $n_0=60.$

If one considers lower $n_0$ (say around $n_0=30$) when the effective
$\hbar$ is bigger, the anharmonic terms in the Hamiltonian become
more important and the spontaneous emission spectrum contains other
lines, at $\Omega+k\omega_-$ with $k\ge 2$ or combinations, e.g.
at $\Omega+2\omega_--\omega_+$. This nicely correlates with our
study of ionization mechanism \cite{zdb98}. Starting around $n_0=40,$
ionization clearly occurs via a chaos assisted tunneling mechanism
(obeying exponential behavior with $\hbar$ change) with clear
deviations from such a behavior for lower $n_0$.

Similarly, much weaker
microwave amplitudes (say $F_0=0.005$) when the resonance island is
so small that the quantum eigenstate does not fit into it,
lead to a similar behavior of the spontaneous emission
spectrum - the presence of overtones.
Still we have checked that in quite a broad range of $0.01 < F_0
<0.06,$ the harmonic approximation works similarly to the case
shown in Fig.~\ref{exdip} and Fig.~\ref{exsp}, provided no
accidental avoided crossing with other Floquet states occurs
\cite{zdb:zfp}.

While the calculations presented in this paper are consistently
done at the lowest order of the harmonic approximation using
Cartesian coordinates, similar calculations can be done using, for
example, the spherical coordinates \cite{keb,ke}. Consistently, the
results at lowest order are of course identical. However, it is known
that -- within the harmonic approximation -- the wavefunction in spherical
coordinates is a better approximation to the ``exact" wavefunction than
the corresponding one in Cartesian coordinates, as it incorporates some
of the anharmonic terms; the price to pay is the breaking of the
selection rules, Eqs.~(\ref{sr1})-(\ref{sr2}). Hence, the transition
rates calculated using harmonic wavefunctions in spherical coordinates
incorporate already some of the higher order terms and might be closer
to the ``exact" results. Note, however that if the calculation is
performed consitently at any order in perturbation theory (the anharmonic
terms being the perturbation), the results should of course be independent
of the choice of coordinates.

\section{Conclusions}

It is worth stressing again that, in the semiclassical limit
(small $\Omega$ or large $n_0$), the dominant contribution to
the spontaneous emission comes from the elastic component
at frequency $\Omega$ which {\it does not} destroy the wave packet
but rather leads to the decoherence in its time-evolution.
The wave packet converts the energy of the driving photon into a
spontaneously emitted photon, moving at the same time on the circular
orbit.

\null From a practical point of view, the spontaneous emission rates
are several orders of magnitude smaller than the ionization rate
in a typical situation $n_0=60$ and $F_0=0.05$ \cite{zdb98},
and thus unlikely to be observable. However, this ratio can be
made smaller by increasing $n_0$. Indeed, as shown in \cite{zdb98},
the typical ionization rate decreases {\em exponentially} with
$n_0$ (because it is due a chaos assisted tunneling process) while
the spontaneous decay rates decrease {\em algebraically} with
$n_0.$ For very large $n_0,$ spontaneous emission will
eventually become the dominant process. For $F_0=0.05$, the
cross-over should take place around $n_0=200$
which makes it very difficult to observe experimentally.
Another interesting possibility is to decrease the scaled
field $F_0$ so that the ionization rate drops considerably.
According to numbers shown in \cite{zdb98}, the cross-over
for $n_0=60$ should take place around $F_0=0.03$
in a regime where the harmonic approximation is still
valid (the results are not shown since they are very similar to those
depicted in Fig.~\ref{exsp}).
 In any case, the total decay rate is of the order of
few tens of photons per second, making the experimental observation
rather difficult.

\section{Acknowledgments}
We thank I.~Bialynicki-Birula and A.~Buchleitner for sending
us preprints of their work prior to publication.
Partial support of KBN  in the form of travel grants
is acknowledged. We are grateful to French Embassy
in Poland for support under
billateral collaboration program 67209.
CPU time has been provided by IDRIS and Pozna\'n SCNC.
Laboratoire Kastler-Brossel, de l'Ecole Normale Sup\'erieure et de
l'Universit\'e Pierre et Marie Curie, is unit\'e associ\'ee 18 du CNRS.


\begin{references}
\bibitem{foot1} We shall not consider typical quantum mechanical effects
important when the number of photons in the occupied mode is small,
such as quantum revivals. The equivalence between dressed state and
Floquet picture is obtained, let us recall, neglecting the difference
between $N$ and $N+1$, where $N$ is the (very large) number of
photons in the occupied mode.
\bibitem{rotframe} In the context of quantum optics of a two-level
system, the rotating wave approximation makes it possible to define
a time-dependent unitary transformation producing a static Hamiltonian.
As there is a formal analogy, this is known as passing to 
the ``rotating frame" although this is in general
not a purely geometrical transformation. Only in the case of a circularly
polarized field do the two acceptions of ``rotating frame" coincide.
\bibitem{foot6} We have chosen the length gauge for the description of the
empty modes of the
electromagnetic field. Of course, one could use a different gauge, such
as the velocity gauge. The matrix elements coming into play are different,
but we checked that the decay rates of the various spontaneous transitions 
are -- as they should -- gauge independent. Similarly, we could use another
gauge to describe the driving microwave field. This produces a different
Floquet Hamiltonian and a different static Hamiltonian in the rotating
frame, hence different creation/annihilation operators for the normal modes.
However, the energy levels of the dressed atom -- and consequently
the frequencies of the normal modes -- do not depend on the choice of gauge.
The same is true for the decay rates. The ``exact" computations
shown in this paper have been performed in velocity gauge.
\bibitem{foot4} The calculation is here done for the density of
electromagnetic modes in free space. The decay rates can be substantially
modified in the presence of a cavity, as could be used in a real
experiment \protect{\cite{vacuum}}.
\bibitem{foot7} In \protect{\cite{bb97}}, the matrix elements of the
dipole operator are calculated using explicit forms of the wavefunctions
in configuration space. We prefer to use no explicit representation
of the eigenstates and manipulate creation/annihilation operators
in the normal modes: extension of the present results to the
decay of ``excited" non dispersive wavepackets is much easier this way.
\bibitem{foot5} Note however that the spontaneous 
decay rate per Kepler period
as given in \protect{\cite{bb97}} at the end of page 3630
has a $2\pi$ factor
wrongly placed. It is thus ``only" 4 orders of magnitude smaller
than the ionization rate.
\end{references}
\end{document}